\documentclass[aps,prd,twocolumn,nofootinbib]{revtex4-2}
\usepackage{epsfig}
\usepackage{graphicx}
\usepackage{dcolumn}
\usepackage{amssymb,amsmath}
\usepackage{mathrsfs}
\begin{document}

\title{Nuclear clock based on the Th V ion}

\author{V. V. Flambaum$^1$}\email{v.flambaum@unsw.edu.au}
\author{V. A. Dzuba$^1$}\email{v.dzuba@unsw.edu.au}
\author{E. Peik$^2$}\email{ekkehard.peik@ptb.de}

\affiliation{$^1$School of Physics, University of New South Wales, Sydney 2052, Australia}
\affiliation{$^2$Physikalisch-Technische Bundesanstalt, 38116 Braunschweig, Germany}

\begin{abstract}

We propose that a nuclear clock based on the Th V ion can surpass the accuracy of clocks built with other thorium ions. The Th$^{4+}$ ion has a rigid closed-shell core with zero total electron angular momentum, suppressing frequency shifts from black-body radiation and stray external fields that act mainly on electrons.
We calculate the energy shift of the nuclear clock transition frequency in $^{229}$Th due to the Coulomb field of atomic electrons and find a relative frequency difference of $2.8 \times 10^{-7}$ between Th IV and Th V - twelve orders of magnitude larger than the projected $10^{-19}$ fractional uncertainty of a nuclear clock.
We also perform calculations for Th V energy levels, ionization potential, static polarizability, and the black-body radiation shift of the nuclear line.
Additionally, we determine the nuclear transition frequencies in two thorium ions and neutral atom:
 $\omega_N=2,020,406.964(70)$ GHz in Th III,  $\omega_N=2,020,408.264(100)$ GHz  in Th II, and 
  $\omega_N=2,020,408.364(100)$ GHz in Th I.

\end{abstract}

\maketitle

\section{Introduction}

The nucleus of the $^{229}$Th isotope has the unique feature of having a very low-energy excited state connected to the ground state by a magnetic dipole (M1) transition (see, e.g. reviews~\cite{Rev,Rev1} and references therein). 
The measurement of the energy of this nuclear clock transition in $^{229}$Th has been an ongoing effort for many years~\cite{wn1,wn2,wn3}.  Recently a very high accuracy has been been obtained 
in Refs.  \cite{Katori,Th-wn0,Th-wn,Th-wn1,Th-wn2}. 
The latest, most precise measurement, gives the value   $\omega_N = 2,020,407,384,335(2)$~kHz (67393 cm$^{-1}$) for $^{229}$Th dopant ions embedded in a calcium fluoride crystal~\cite{Th-wn1}. A still higher accuracy is expected for a nuclear clock with trapped ions \cite{PeikTamm,Th3+}, because of the absence of the crystal electric and magnetic fields and because of the ultralow temperatures that are attainable with laser cooling.  

This low-energy nuclear transition  attracted many researchers planning to  build a nuclear clock of exceptionally high accuracy  - see e.g. ~\cite{PeikTamm,Thm}.
The relative uncertainty is expected to reach $10^{-19}$~\cite{Th3+}. In addition, there are strong arguments that this nuclear clock would be very sensitive to physics beyond the standard model, including space-time variation of the fundamental constants, violation of Lorentz invariance and the Einstein equivalence principle, and search  for scalar and axion dark matter fields ~\cite{vara,Lorentz,varq,vara1,vara2,vara3,vara4,Arvanitaki,Stadnik,Gilad}.
There are plans to use trapped Th ions ~\cite{Th3+,Th+,dr} and  solid-state Th nuclear clocks~\cite{Hudson,ThSS,ThSS1}.
It was shown in Ref.~\cite{dwn} that  in all these systems the frequency of the nuclear clock will be different. 
This is due to the Coulomb interaction of atomic electrons with the nucleus, leading to the significant 
electronic shift of the nuclear transition frequency. Additionally, there is a frequency  shift due to magnetic interaction between atomic electrons and the nucleus. 
Specifically, the electronic shifts in  Th~IV, Th~III, Th~II ions and neutral Th atom have been calculated  in Ref.~\cite{dwn} (in the present paper we calculated these shifts with a different method which provides a higher accuracy). There may also be a shift due to the interaction with electric and magnetic fields  in the solid or with external fields for the ion trap. 

In this paper we consider a nuclear clock based on the Th~V ion.  
This ion has zero electron angular momentum in the ground state, effectively eliminating magnetic interaction effects for the electron shell, including the hyperfine structure in the nuclear clock transition. Coupling of electron and nuclear angular  momenta  into a hyperfine structure component in other ions could produce an entanglement between electron and nuclear  variables. There is no such entanglement in the Th V ion.     
Since electrons interact with black-body radiation and stray fields many orders of magnitude stronger than the nucleus, eliminating  effects of this interaction on the nucleus  helps to improve the stability and accuracy of the nuclear clock (see below).

Since Th~V does not possess an optical electronic resonance in the wavelength range of readily available lasers, the readout of the nuclear state can be performed via a co-trapped quantum logic ion that will also provide sympathetic laser cooling \cite{Schmidt:2005}. 
Either Ca~II or Sr~II can be used  as a logic ion.
These  ions are routinely laser-cooled in many laboratories around the world (see~\cite{Ludlow} and references therein). 
The charge-to-mass ratios $q/m$  are relatively close for these ions: 0.0175 for  $^{229}$Th~V,  0.0208 for  $^{48}$Ca~II and 0.0114 for  $^{88}$Sr~II (in units of $e/$amu).

In the present work we discuss benefits of using Th~V ion for a nuclear clock, consider the effect of electrons on the nuclear transition  frequency, including  the change of the frequency between Th~V and other Th ions.  Additionally, we calculate energy levels of Th~V.
Electronic structure of the Th~V ion was studied theoretically in Ref.~\cite{LaIV}. This included calculations of energy levels for the low even states and M1 and E2 transition amplitudes.
We perform calculations of energy levels for both  even and odd states, ionisation potential and the effect of electrons on the nuclear clock frequency. Experimentally known energy levels of the La~IV ion are used to control the accuracy of the calculations in the present work and in Ref.  \cite{LaIV}.

\section{The effect of electrons on the nuclear transition frequency.}

Electrons influence the nuclear transition frequency in two distinct ways. The first type of effects introduces uncertainties in frequency measurements, while the second results in a frequency shift between systems with different electron configurations~\cite{dwn}.

Two classes of electronic states have been identified in previous work as particularly well suited for the interrogation of the nuclear resonance: Spherically symmetric states with $J=0$ or $J=1/2$~\cite{PeikTamm}, and stretched states with $F=J+I$~\cite{Th3+}. In the Th~IV ion, the first condition is fulfilled by a metastable $7s~^2S_{1/2}$ state, and the second, more generally usable, also for the ground state.     

 A detailed analysis of uncertainties in the Th~IV ion was presented in our previous work ~\cite{Th3+}.  Later Beloy, in Ref.~\cite{Beloy}, argues that the dominant systematic effect in Th~IV is the second-order ac Zeeman shift, which arises from the mixing of hyperfine structure levels due to the ac magnetic field in a Paul trap. However, most of these effects are either absent or significantly suppressed in the Th~V ion due to its closed-shell configuration with zero total electron angular momentum.
The Th~V ion has no hyperfine structure in its ground state, and its nuclear magnetic moment, nuclear electric quadrupole moment, and nuclear polarizability are very small  on the atomic scale. 

Additional suppression of some  systematic frequency shifts is due to  the  shielding  of the nucleus from external electric fields  by atomic electrons - see e.g.  Ref. \cite{shielding}. For a static electric field or a  low frequency field (when the frequency is smaller than electron excitation energies),  the shielding factor is $Z_i/Z=4/90$, where $Z_i=4$ is the ion charge, and $Z=90$ is the nuclear charge. This  suppression factor also applies for the blackbody radiation (BBR)  electric field  acting on the  nucleus. However, the main reason  for BBR shift suppression for a nuclear transition is the very small nuclear polarizability, which is typically 15  orders of magnitude smaller than atomic polarizability   ( the BBR shift is proportional to the polarizability difference between ground and excited clock states).   A suppression of the effects of collisions  between ions is also expected since the electric field of another ion is suppressed at the nucleus.

The electron cloud polarizability  becomes relevant  since  the electron state is affected by the nuclear transition. Indeed, the nuclear spin, magnetic moment and electric quadrupole moment are different in the ground and excited nuclear states \cite{Thm,Th-wn1}. Their effect appears due to mixing of electron states by the hyperfine interaction (hfi)  with the nucleus which affects the electron polarizability.  The energy denominators in the formula  for the polarizability are also affected by hfi.  These effects are suppressed by the ratio of the hfi matrix element   to the electron excitation energy, which is $\sim 10^{-5}$. Moreover, in Th V, the  first order hfi effect vanishes since  electron angular momentum is zero.  

 However, there is an effect of the nuclear radius change  $\delta \langle r^2 \rangle \approx 0.010$ fm$^2$ (see discussion of data for $\delta \langle r^2 \rangle$ below).  According to our calculation, static polarizability of  Th V is $\alpha_0=7.78 a_B^3$, where $a_B$ is the Bohr radius. The difference between polarizabilities due to the change in the nuclear  radius  is  $\delta \alpha_0 = -0.95  \times 10^{-4}  a_B^3 \delta \langle r^2 \rangle/$fm$^2= -1.0 \times 10^{-6}  a_B^3$. This gives a BBR shift  of the nuclear transition at the temperature T=300 K  equal to $\delta \nu=8.6  \times 10^{-9}$ Hz, i.e.  $\delta \nu/\nu =4.3   \times 10^{-24}$.

Because the nuclear spin of $^{229}$Th is a half-integer ($I=5/2$ in the nuclear ground state and $I'=3/2$ for the isomer) a small first-order frequency shift with the magnetic field strength is present for all Zeeman components of the nuclear transition in the Th~V electronic ground state. The effect can be corrected for by averaging the frequency under constant magnetic field strength over two symmetric components of the Zeeman multiplet, for example the pair of resonances originating from the $m_F=\pm1/2$ levels under excitation with circularly polarized light with a sensitivity of $\pm0.04$\,Hz$/\mu$T. The small sensitivity results from the near equal absolute values of the $g$-factors of ground state and isomer with opposite sign \cite{Thm,Porsev2021}.
A similar method is routinely applied in optical clocks, for example with $^{87}$Sr and $^{87}$Sr$^+$~\cite{Ludlow}. 

The most notable remaining systematic frequency shift in the $^{229}$Th nuclear clock is the relativistic second-order Doppler shift due to micromotion in the ion trap. 
It was demonstrated in \cite{Th3+,micromotion} that this shift is small and can be controlled on the level ($\Delta \nu/\nu \sim 10^{-20}$) in a well designed ion trap.

We now consider the frequency shift between Th~IV and Th~V ions.
It has been demonstrated in our previous work~\cite{dwn}  that  the Coulomb interaction of atomic electrons with the nucleus leads to a significant 
shift of the nuclear transition frequency. 
This shift for electronic state $a$ is given by
\begin{equation}\label{e:IS}
\Delta E_a = F_a \delta \langle r^2 \rangle,
\end{equation}
where $F_a$ is the field shift constant of state $a$ which can be obtained from atomic calculations; $\delta \langle r^2 \rangle$ is the change of the nuclear root-mean square radius between the excited and ground nuclear states. We use the  value for $\delta \langle r^2 \rangle$  derived in Ref.~\cite{dr}, 
$^{229m,229}\delta \langle r^2 \rangle=0.0105(13)~{\rm fm}^2$. 
Thus, the difference of the nuclear frequencies between Th~IV and Th~V is given by 
\begin{equation}\label{e:ISN}
\Delta \omega_N = (F_a (\text{Th~IV}) -  F_a(\text{Th~V}) ) \delta \langle r^2 \rangle,
\end{equation}
State $a$ is the ground electronic state of the ion.

In our previous work~\cite{dwn}, we calculated the field shift constant $F_a$ for Th~I, Th~II, Th~III, and Th~IV. The closed-shell core of all these ions is the same, corresponding to the Th~V ion. In this approximation, the valence electron contribution to Th~V is effectively $F_a(\rm Th~V) \equiv 0$, allowing the result for Th~IV from Ref.~\cite{dwn} to be used in calculating the frequency shift via Eq.(\ref{e:ISN}).

Note that the modification of the electronic core contribution due to the additional valence $5f$ electron in Th~IV has been included as a core relaxation effect in \cite{dwn}. To improve reliability and estimate the accuracy of our results, we employ two additional methods to calculate the energy shift in Eq.(\ref{e:ISN}). In these approaches, the $6p$ electrons are treated as part of the valence space (see Appendix), meaning that the field shift constant for Th~V is no longer zero.

First, we perform calculations using two different values of the nuclear radius and apply Eq.~(\ref{e:IS}) to determine the field shift constant $F_a$.

Second, we treat the nuclear radius change as a perturbation and employ the random-phase approximation (RPA), which incorporates core relaxation effects, to compute the field shift constant $F$ (see Appendix for details).

The results are presented in Table~\ref{t:F1}. 
We also added the results for other Th ions obtained via change of the nuclear radius, since this method includes corrections missing in the RPA method used in Ref.~\cite{dwn}, and is expected to be significantly more accurate (see Appendix). 

The agreement between the values of $F$ for Th~IV and Th~V obtained using two different methods demonstrates a correlation between the complexity of the system and the accuracy of the results. The greater the number of valence electrons, the larger the observed difference.
 The values of $\Delta F$ are further affected by cancellations in case when the values of $F$ for two ions are close.
Assuming that the values obtained with the $\Delta R_N$ method are more accurate, we use the value $\Delta F= -54.2$~GHz/fm$^2$ for the Th IV - Th V pair. 

Taking this value and the change of nuclear radius between the ground and isomeric nuclear states
$\delta \langle r^2 \rangle=0.0105(13)~{\rm fm}^2$, we obtain the nuclear frequency shift between Th~IV and Th~V ions to be -570~MHz = $-2.4\times 10^{-6}$~eV. The relative shift $|\Delta \omega_N/\omega_N| = 2.8 \times 10^{-7}$ is larger than projected uncertainty of the frequency measurement by twelve orders of magnitude. 
See Table~\ref{t:dw} for comparison with other ions.

\begin{table}
\caption{\label{t:F1}Field shift constants $F$ and their differences $\Delta F$ (GHz/fm$^2$) for the ground states of Th~I, Th~III, Th~IV and Th~V ions found in different approaches: first, by changing nuclear radius ($\Delta R_N$) and second in RPA calculations. 
Two more approaches were used for Th~IV: it was treated as a system with one external electron and as a system with seven valence electrons ($N_v$). }
\begin{ruledtabular}
\begin{tabular}   {llrr}
\multicolumn{1}{c}{$N_v$}&
\multicolumn{1}{c}{Ions}&
\multicolumn{1}{c}{$\Delta R_N$}&
\multicolumn{1}{c}{RPA}\\
\hline
3 &$F$(Th~II)  &   51.7  & 49.6\footnotemark[1]     \\
2 &$F$(Th~III) &   -68.2 & -68.0\footnotemark[1]        \\
1 &$F$(Th~IV)  &   -54.2  & -55.0\footnotemark[1]       \\
7 &$F$(Th~IV)  &   -269  & -281      \\
6 &$F$(Th~V) &   -222  & -230       \\
3-2 & $\Delta F$(Th~II - Th~III) & 119.9 & 117.6 \\
2-1 & $\Delta F$(Th~III - Th~IV) &  -14.0 & -13.0 \\
1-0 & $\Delta F$(Th~IV - Th~V) &  -54.2 & -55.0\footnotemark[1]  \\
7-6 & $\Delta F$(Th~IV - Th~V) &  -47 & -51 \\
\end{tabular}
\footnotetext[1]{Ref.~\cite{dwn}.} 
\end{ruledtabular}
\end{table}


\begin{table}
  \caption{\label{t:dw}   
   Shift in the nuclear transition frequency $\omega_N$ between different ions of $^{229}$Th. Results for Th~I to Th~IV and the bare $^{229}$Th nucleus are taken from Ref.~\cite{dwn}, while the frequency difference between Th~IV and Th~V is calculated in the present work.}  

\begin{ruledtabular}
\begin{tabular}   {lcl crc}
\multicolumn{3}{c}{Ions}&
\multicolumn{2}{c}{$\Delta \omega_N$}&
\multicolumn{1}{c}{$\Delta \omega_N/\omega_N$}\\
&&&\multicolumn{1}{c}{(GHz)}&
\multicolumn{1}{c}{(eV)}& \\
\hline
Th I  &$-$& Th II   &  0.10  & $4.1 \times 10^{-7}$ &   $5.0 \times 10^{-8}$ \\
Th II  &$-$& Th III & 1.3     &  $5.4 \times 10^{-6}$ &    $6.4 \times 10^{-7}$ \\
Th III &$-$& Th IV & -0.15  &   $-6.2 \times 10^{-7}$ &   $-7.4 \times 10^{-8}$ \\
Th IV &$-$& Th V & -0.57  &   $-2.4 \times 10^{-6}$ &   $ -2.8 \times 10^{-7}$ \\ 
Th V &$-$& Th XCI\footnotemark[1] & $2.1 \times 10^{4}$  &   $8.7 \times 10^{-2}$ &   $ 1.0 \times 10^{-2}$ \\ 
\end{tabular}			
\footnotetext[1]{Bare $^{229}$Th nucleus.}
\end{ruledtabular}
\end{table}

 After a preprint of the present paper  \cite{present} appeared in arxiv, another  preprint \cite{Perera} appeared with extended calculations of the electron-induced  energy shifts of the nuclear transition frequency to Th ions in the solid state environment. Combined  with the measurements  ~\cite{Th-wn1} and calculations of the frequency  difference between the free ion and Th ion in solid,  
 this allowed them to find the absolute values of the nuclear transition frequencies $\omega_N=2,020,407.114(70)$ GHz  in  Th IV and  $\omega_N=2,020,407.648(70)$ GHz  in Th V ions.  The difference of these frequencies -0.534 GHz agrees with our result -0.570 GHz. An apparent  deviation   0.036 GHz  (which is  much smaller than the estimated uncertainty of 0.1 GHz) is  due to the slightly different value of $\delta \langle r^2 \rangle=0.0097(26)$fm$^2$  \cite{Katori} used in Ref.  \cite{Perera} (we used  
$^{229m,229}\delta \langle r^2 \rangle=0.0105(13)~{\rm fm}^2$ \cite{dr}). The experimental  uncertainty in $\delta \langle r^2 \rangle$ at the moment is the main source of errors in  the calculated values of  $\omega_N$. Improved values of  $\delta \langle r^2 \rangle$  and calculated values of the field shift constants $F$ in Table  \ref{t:F1} will allow one to improve predictions of  $\omega_N$. 
 
  Using our calculations of  the frequency differences between Th ions  from Table \ref{t:dw} and the  frequencies in Th IV and Th V from Ref. \cite{Perera},  we obtain $\omega_N=2,020,406.964(70)$ GHz in Th III,  $\omega_N=2,020,408.264(100)$ GHz  in Th II, 
  $\omega_N=2,020,408.364(100)$ GHz in Th I, and  $\omega_N=1,999,700(3000)$ GHz = 8.270(12) eV for the  bare nucleus  $^{299}$Th$^{+90}$. The result for the latter is  $\omega_N=$8.272(22) eV  in Ref. \cite{Perera}.  

\section{Energy levels of Th~V and La~IV}

 To  provide theoretical support for further experiments with Th ions, it may be useful to calculate unknown  electronic energy levels  of Th~V  ion.  
The Th~V and La~IV ions have similar electronic structures, both featuring closed shells with an outermost subshell configuration of $ns^2np^6$ in the core ($n=5$ for La~IV and $n=6$ for Th~V). Experimental data for the La~IV spectrum is available in the NIST database~\cite{NIST}, whereas no experimental data exist for Th~V. Therefore, we estimate the accuracy of our method by comparing our calculated energy levels for La~IV with the available experimental data.

Calculations of several low-energy even-parity levels in La~IV and Th~V were performed in Ref.~\cite{LaIV} using an advanced method that combines configuration interaction (CI) with a linearized coupled-cluster approach~\cite{CI+SD}. A sophisticated procedure was employed to select the most important configurations, resulting in agreement with measured  La~IV levels within 1\%.

In the present work, we calculate a significantly larger number of levels. 
We use a less computationally demanding but highly efficient method that combines CI with perturbation theory, known as the CIPT method~\cite{CIPT} (see Appendix for details). The efficiency is gained by reducing the size of the effective CI matrix, which is generated using many-body perturbation theory. 

The results for fifteen lowest excited states of La~IV are presented in Table~\ref{t:EL}. 
Comparison with experiment shows satisfactory agreement for even states ($\sim$ 6\%) and very good agreement for odd states ($\sim$ 1\%). 

Excitation energies for the Th~V ion are 
well outside of the optical region (e.g. 61 nm wavelength for the resonance line to the $6d~1^-$~state). 
This means that using the electronic bridge process~(see, e.g. \cite{EB1,EB2,EB3}) for nuclear excitation is hardly possible. 
Note however that some energy intervals between excited states of Th~V come close to the nuclear excitation energy $\omega_N$ =67353~cm$^{-1}$. 
For example, if we take the first excited state, $E=138922 ~{\rm cm}^{-1}$ and add to it the value of the nuclear frequency, we get $E=206275 ~{\rm cm}^{-1}$, which is in the area where the spectrum is dense and a resonance is possible.

The ionization potential of Th~V was calculated as the energy difference between ground states of Th~V and Th~VI. It is in good agreement with previous results of Refs.~\cite{NIST,Rodrigues}.

\begin{table}\label{t:EL}
  \caption{Experimental~\cite{NIST} and calculated excitation energies (in cm$^{-1}$) and $g$-factors of
  the La~IV ion. Ground state configuration is [Pd]$5s^25p^6$, symbols F and D indicate leading configurations for outermost electrons:
  symbol F indicates the $5p^54f$ configuration, symbol D indicates the $6p^55d$ configuration.The symbol
$J^p$ stands for the total angular momentum and parity.}
\begin{ruledtabular}
\begin{tabular}   {rl  lc rl rr}
\multicolumn{3}{c}{State}&
\multicolumn{1}{c}{$J^p$}&
\multicolumn{2}{c}{This work}&
\multicolumn{1}{c}{NIST\cite{NIST}}&
\multicolumn{1}{c}{Ref.~\cite{LaIV}}\\
\hline
1 &  GS &$^1$S      &  0$^+$  &        0 &  0.0000 &       0.0 &  0 \\
2 &  F  &2[3/2] &  1$^+$   &   151189 &  0.5000 &  143354.7 & 142168   \\
3 &  F  &        &  2$^+$   &   153213 &  1.1373 &  145949.0 & 144910   \\
4 &  F  &2[5/2]  &  3$^+$   &   157704 &  1.2750 &  149927.1 & 149153   \\
5 &  F  &        &  2$^+$   &   169914 &  0.8320 &  160486.4 & 160592  \\
6 &  F  &2[7/2]  &  4$^+$   &   158020 &  1.0408 &           & 157252  \\
7 &  F  &        &  3$^+$   &   162326 &  0.8803 &  153339.1 & 153130   \\
8 &  D  &2[1/2] &  0$^-$   &   156068 &  0.0000 &  156100.3 &   \\
9 &  D  &        &  1$^-$   &   157993 &  1.4444 &  158412.6 &   \\
10 &  D  &2[3/2] &  2$^-$   &   162794 &  1.4185 &  162867.6 &   \\
11 &  D  &        &  1$^-$   &   182319 &  0.5782 &  181155.0 &   \\
12 &  D  &2[7/2] &  4$^-$   &   163680 &  1.2500 &  163693.3 &   \\
13 &  D  &        &  3$^-$   &   165503 &  1.0933 &  165070.7 &   \\
14 &  D  &2[5/2] &  2$^-$   &   168593 &  0.8897 &  167921.7 &   \\
15 &  D  &        &  3$^-$   &   173718 &  1.2041 &  173335.5 &   \\
\end{tabular}
\end{ruledtabular}
\end{table}

\begin{table}\label{t:ELTh}
  \caption{Calculated excitation energies (in cm$^{-1}$) and $g$-factors of
  the Th~V ion. Ground state configuration of outermost electrons is $6s^26p^6$, symbols F, D, P  and S indicate leading configurations for outermost electrons: symbol F indicates the $6s^26p^55f$ configuration, symbol D indicates the $6s^26p^56d$ configuration,
  symbol P indicates the $6s^26p^57p$ configuration, symbol S indicates the $6s^26p^57s$ configuration.
  The symbol $J^p$ stands for the total angular momentum and parity. IP stands for ionisation potential.}
\begin{ruledtabular}
\begin{tabular}   {rl  c rl r}
\multicolumn{2}{c}{State}&
\multicolumn{1}{c}{$J^p$}&
\multicolumn{2}{c}{This work}&
\multicolumn{1}{c}{Ref.~\cite{LaIV}}\\
\hline
1 &   GS & 0$^+$ &        0 &  0.0000 & 0 \\
  2 &   F  & 1$^+$ &   138922 &  0.5000 & 134995 \\
  3 &   F  & 2$^+$ &   143129 &  1.0922 & 139842 \\
  4 &   F  & 4$^+$ &   146618 &  1.0395 & 143160  \\
  5 &   F  & 5$^+$ &   147335 &  1.2000 & 144714 \\
  6 &   F  & 3$^+$ &   149812 &  1.1985 & 146314 \\
  7 &   F  & 3$^+$ &   151891 &  0.9574 & 148081  \\
  8 &   F  & 4$^+$ &   157311 &  1.1657 & 154552  \\
  9 &   D  & 0$^-$ &   159866 &  0.0000 &  \\
 10 &   F  & 2$^+$ &   161992 &  0.8633 & 159248 \\
 11 &   D  & 1$^-$ &   163431 &  1.3538 &  \\
 12 &   D  & 3$^-$ &   169790 &  1.0863 &  \\
 13 &   D  & 2$^-$ &   170331 &  1.2633 &  \\
 14 &   D  & 4$^-$ &   171779 &  1.2500 &  \\
 15 &   D  & 2$^-$ &   174617 &  1.0379 &  \\
 16 &   D  & 3$^-$ &   181285 &  1.2157 &  \\
 17 &   D  & 1$^-$ &   195273 &  0.8926 &  \\
 18 &   D  & 2$^-$ &   196523 &  1.4928 &  \\
 19 &   D  & 1$^-$ &   201481 &  1.0393 &  \\
 20 &   F  & 3$^+$ &   208580 &  0.8534 & 205597  \\
 21 &   F  & 3$^+$ &   212757 &  1.1575 & 210417 \\
 22 &   F  & 4$^+$ &   215662 &  1.0948 & 214015 \\
 23 &   F  & 2$^+$ &   219266 &  0.8780 &  \\
 24 &   D  & 2$^-$ &   233093 &  0.7864 &  \\
 25 &   P  & 1$^+$ &   238044 &  1.6338 &  \\
 26 &   P  & 2$^+$ &   239200 &  1.1530 &  \\
 27 &   D  & 3$^-$ &   243125 &  1.1146 &  \\
 28 &   P  & 1$^+$ &   254730 &  1.2264 &  \\
 29 &   D  & 1$^-$ &   254896 &  0.8935 &  \\
 30 &   S  & 0$^-$ &   263152 &  0.0000 &  \\
\hline
IP   &       & 3/2$^-$ & 472331\footnotemark[1] & \\
\end{tabular}
\footnotetext[1]{Other theory, IP = 468000(15000)~\cite{NIST,Rodrigues}.}
\end{ruledtabular}
\end{table}

%


This work was supported by the Australian Research Council Grant No. DP230101058.
EP acknowledges support from the European Research Council (ERC) under the European Union's  Horizon 2020 research and innovation programme (Grant Agreement No. 856415), the Deutsche Forschungsgemeinschaft (DFG)  - SFB 1227 - Project-ID 274200144 (Project B04), and by the Max-Planck-RIKEN-PTB-Center for Time, Constants and Fundamental Symmetries.
\appendix 
\section{Method of calculation}

In the present work we use an efficient method which combines CI with perturbation theory (the CIPT method~\cite{CIPT}). 

We perform the calculations in the $V^{N-1}$ approximation which means that the initial Hartree-Fock (HF) procedure is done for an ion with one hole in the outermost subshell ($5p^5$ for La~IV and $6p^5$ for Th~V).
The single-electron basis states are calculated in the field of frozen core using the B-spline technique~\cite{B-spline}. 

It was demonstrated in Ref.~\cite{LaIV} that the $5s$ electrons in La~IV and $6s$ electrons in Th~V should be attributed to valence space for better results.
Any excited states of the ions involve at least one excitation of an electron from either the $ns^2$ or $np^6$ subshell, making the ion an open-shell system with eight electrons in the open shells. Full-scale configuration interaction (CI) calculations are challenging for eight valence electrons.
Therefore, we use a simplified but very efficient approach especially developed for open-shell systems.
It is called the CI with perturbation theory (CIPT) method~\cite{CIPT}. 
The CIPT equations have the form
\begin{equation}
\langle i|H^{\rm eff}|j\rangle = \langle i|H^{\rm CI}|j\rangle+\sum_{k}\frac{\langle i|H^{\rm CI}|k\rangle\langle k|H^{\rm CI}|j\rangle}{E-E_k}. 
\label{e:CIPT}
\end{equation}
Here $H^{\rm CI}$ is the CI Hamiltonian for eight external electrons
\begin{equation}
H^{\rm CI} = \sum_n^8 \hat H^{\rm HF}_n +\sum_{n<m}^8 \frac{e^2}{r_{nm}},
\label{e:HCI}
\end{equation}
$\hat H^{\rm HF}_n$ is the relativistic Hartree-Fock (HF) operator for the valence electron number $n$. 
Indices $i,j,k$ in (\ref{e:CIPT}) numerate eight-electron basis states which are constructed by exciting one or two electrons from the reference $ns^2np^6$ configuration.
All basis states are divided on the energy scale into two groups, low-energy states ($i,j$), and high-energy states ($k$). Low-energy states are included directly into the CI matrix while high-energy states are included perturbatively.
 $E$ in (\ref{e:CIPT}) refers to the energy of the state of interest, and $E_k$ is the diagonal matrix element for high-energy states, $E_k =\langle k| H^{\rm CI}| k\rangle$. Summation in (\ref{e:CIPT}) goes over all high-energy states. 
The energies $E$ and wave functions $X$ are found by solving the matrix eigenvalue problem
\begin{equation}
\left(H^{\rm eff}-E{I}\right) X = 0,
\label{e:CI} 
\end{equation}
with $H^{\rm eff}$ matrix given by (\ref{e:CIPT}), while $I$ is the unit matrix. Note that the $H^{\rm eff}$ matrix depends on unknown energy $E$.
Therefore, iterations over energy are needed. Usually five to ten iterations are enough for full convergence.

The results for fifteen lowest excited states of La~IV are presented in Table~\ref{t:EL}. 
Comparison with experiment shows satisfactory agreement for even states ($\sim$ 6\%) and very good agreement for odd states ($\sim$ 1\%). 

Calculations of the field shift constant on the first stage are very similar to the calculation of energy levels. One significant difference is that the $6s$ electrons are moved to the core and the CI equations are solved for six ($6p^6$) or seven ($6p^65f$) valence electrons. This is because the $6s$ electrons give large contribution to the field shift constants of both ions, Th~IV and Th~V, enhancing numerical error in the difference $\Delta F$. Moving them to the core leads to more stable results.

We perform the calculations in two different ways. First, we perform the calculations with two values of nuclear radius, then apply (\ref{e:IS}) to find the values of $F_a$.
This way is simpler  but  it may be  sensitive to numerical noise. After changing the nuclear radius one needs to perform extra HF iterations to account for the core relaxation effect. These iterations themself produce some shift in energy which might be comparable to the isotope  shift. For this reason the method does not work for light atoms since the shift of energy is very small there. However, this method is sufficiently accurate for heavy ions like Th~IV and Th~V. 
It is important to make an appropriate choice for the value of the change of nuclear radius. It should be large enough to suppress numerical noise in the energy difference. On the other hand, it should not be too large to avoid contributions from higher-order terms,  ($ \sim \delta \langle r^4 \rangle,  \ \delta \langle r^2 \rangle^2$, etc.). We found that $\delta \langle r^2 \rangle \approx 1$ fm$^2$ is sufficently good for these purposes (see Table \ref{t:Fdf}).

\begin{table}
\caption{\label{t:Fdf}Calculation of the field shift constant $F$ for the ground state of Th~IV by changing nuclear radius.}
\begin{ruledtabular}
\begin{tabular}{ccc ccc}
\multicolumn{1}{c}{$R_N$}&
\multicolumn{1}{c}{RMS}&
\multicolumn{1}{c}{$\delta \langle r^2 \rangle$}&
\multicolumn{1}{c}{$E(5f_{5/2})$}&
\multicolumn{1}{c}{$\Delta E$}&
\multicolumn{1}{c}{$F$}\\

\multicolumn{1}{c}{fm}&
\multicolumn{1}{c}{fm}&
\multicolumn{1}{c}{fm$^2$}&
\multicolumn{1}{c}{cm$^{-1}$}&
\multicolumn{1}{c}{cm$^{-1}$}&
\multicolumn{1}{c}{GHz/fm$^2$}\\
\hline
6.9000 & 5.68780  & 0         & -238643.210  & 0            &   0 \\ 
7.0000 & 5.76065 & 0.834 & -238644.719 & -1.509 & -54.24 \\
7.1000 & 5.83362 & 1.680 & -238646.248 & -3.038 & -54.21 \\
7.2000 & 5.90670 & 2.538 & -238647.789 & -4.578 & -54.08 \\
\end{tabular}
\end{ruledtabular}
\end{table}

In a second approach we use the random-phase approximation (RPA)  method to perform the calculations. 
The RPA equations have a form (see e.g. ~\cite{TDHF})
\begin{equation}\label{e:RPA}
(\hat H^{\rm HF} - \epsilon_c)\delta \psi_c = - (\hat F + \delta V_{\rm core}),
\end{equation}
where $H^{\rm HF}$ is the relativistic Hartree-Fock operator for the atomic core, index $c$ numerates single-electron states in the core, $\psi_c$ and 
$\delta \psi_c$ are corresponding single-electron functions and corrections due to the field shift operator $\hat F$, and $\delta V_{\rm core}$ is the change of the self-consistent Hartree-Fock potential due to the change in all core functions. Solving Eqs. (\ref{e:RPA}) self-consistently allows to determine  $\delta V_{\rm core}$. The field shift constant is given by
\begin{equation}\label{e:F}
F_a = \langle a|\hat F + \delta V_{\rm core} |a\rangle.
\end{equation}

We use hat to distinguish between the field shift constant $F$ and the field shift operator $\hat F = \delta V_{\rm nuc}/\delta \langle r^2 \rangle$.
The wave function $|a\rangle$ in (\ref{e:F}) is the many-electron wave function for valence electrons found in the CIPT calculations. 
The RPA equations are linear in $ \delta \langle r^2 \rangle$ by definition; they are also free from numerical noise
caused by extra Hartree-Fock iterations.
Note however that Eq.~(\ref{e:F}) does not take into account some minor contributions like change of the correlation operator $\hat \Sigma$, renormalisation of the wave function, etc. In the end, the RPA method is simpler if minor contributions are ignored. However, with the proper choice of the value for the change of nuclear radius (see, e.g. Table~\ref{t:Fdf}) the alternative  approach is likely to be significantly more accurate since it includes corrections missing in the RPA method.  

\end{document}